\documentclass[english,aps,prl,reprint]{revtex4-1}
\usepackage[T1]{fontenc}
\usepackage[latin9]{inputenc}
\usepackage{bm}
\usepackage{amsmath}
\usepackage{amssymb}
\usepackage{graphicx}

\makeatletter
\usepackage{variables,mathtools}
\usepackage{xr}
\makeatother

\usepackage{babel}
\begin{document}
\title{Off-shell effective energy theory: a unified treatment of the Hubbard
model from $d=1$ to $d=\infty$}
\author{Zhengqian Cheng and Chris A. Marianetti}
\affiliation{Department of Applied Physics and Applied Mathematics, Columbia University,
New York, NY 10027}
\date{\today}
\begin{abstract}
Here we propose an exact formalism, off-shell effective energy theory
(OET), which provides a thermodynamic description of a generic quantum
Hamiltonian. The OET is based on a partitioning of the Hamiltonian
and a corresponding density matrix ansatz constructed from an off-shell
extension of the equilibrium density matrix; and there are dual realizations
based on a given partitioning. To approximate OET, we introduce the
central point expansion (CPE), which is an expansion of the density
matrix ansatz, and we renormalize the CPE using a standard expansion
of the ground state energy. We showcase the OET for the one band Hubbard
model in $d$=1, 2, and $\infty$, using a partitioning between kinetic
and potential energy, yielding two realizations denoted as $\KKF$
and $\XXF$. OET shows favorable agreement with exact or state-of-the-art
results over all parameter space, and has a negligible computational cost.
Physically, $\KKF$ describes the Fermi liquid, while $\XXF$ gives
an analogous description of both the Luttinger liquid and the Mott
insulator. Our approach should find broad applicability in lattice
model Hamiltonians, in addition to real materials systems.
\end{abstract}
\maketitle
Computing the ground state properties of quantum Hamiltonians requires
the search of an exponentially large space of wave functions. To formally
resolve the issue of large dimensionality, one can use effective energy
approaches; which partition the Hamiltonian of a given class into
some external and internal components. The constrained search\cite{Levy19796062}
can then be used to define the energy of the internal components in
terms of expectation values of the external observables and the internal
coupling constants. For example, in density functional theory (DFT)\cite{Hohenberg1964864,Kohn19651133,Jones1989689},
the internal components are the kinetic and interaction energy, and
the external component is the coupling between the density and the
external potential; and the resulting energy functional depends on
the density and the coupling constants of the kinetic and interaction
components. The ground state wave function is then fully determined
from the corresponding external observables and internal couplings,
but such a construction is only useful if robust approximations can
be formulated. 

Here we introduce off-shell effective energy theory (OET), which employs
a wave function ansatz determined from the internal coupling constants
and \textit{both} the internal and external observables. Unlike the
usual effective energy theories, such as DFT, an arbitrary set of
observables will not generally correspond to any ground state within
the class of Hamiltonians; but OET will yield the exact ground state
when minimizing the total energy over the observables. OET opens a
new avenue for developing novel approximations. We introduce the central
point expansion (CPE), which is an expansion of the OET ansatz in
terms of the internal couplings and the internal observables, while
treating the external observables non-perturbatively. The CPE can
then be renormalized (RCPE) using the standard expansion of the energy
in terms of the external observables. Finally, we exploit the possibility
of inverting the role of internal and external components, yielding
a dual formulation of our theory; which will be critical for an accurate
description of the Hamiltonian over all parameter space. 

We apply OET to the single band Hubbard model, which is a canonical
model of interacting Fermions\cite{Gebhard20103642082637,Mahan20000306463385}
with many practical applications, and this will provide a stringent
benchmark of the OET within RCPE. For $d$=1, the Bethe Ansatz (BA) efficiently
provides the exact solution\cite{Lieb19681445,Lieb20031}; while for
$d$=$\infty$, dynamical mean-field theory (DMFT)\cite{Georges199613,Kotliar200453,Vollhardt20121}
provides the solution using numerically exact, but computationally
intensive methods\cite{Kotliar2006865,Gull2011349}. For an arbitrary
dimension, there are powerful but expensive methods which might provide
reliable solutions, though each typically has severe limitations (e.g.
quantum Monte-Carlo\cite{Gubernatis20161107006422,Foulkes200133}
has the minus sign problem\cite{Troyer2005170201,Li2019337}, etc).
Our approach yields favorable agreement with the aforementioned approaches
over all parameter space for the single band Hubbard model in $d$=1,
2, and $\infty$, which is remarkable for a single formalism.

We begin by considering an arbitrary Hamiltonian which has been partitioned
into two parts, $\Ham=\kk[c][][]\kk[o][][]+\xx[c][][]\xx[o][][]$,
where each contribution can be exactly solved. Though this is not
the most general scenario that we consider, it illustrates all key
features of the theory. We begin by choosing $\kk[c][][]\kk[o][][]$
as the internal component and $\xx[c][][]\xx[o][][]$ as the external
component; and this choice is referred to as the $\KKF$ formulation.
The effective energy theory then yields the the density matrix at
a given temperature as 
\begin{equation}
\rho(\kk[c][][],\xx[e][][])=\argmin_{\dm[g]}\{\expt[\kk[c][][]\kk[o][][]+\beta^{-1}\ln\dm[g]][\dm[g]]|\expt[\xx[o][][]][\dm[g]]=\xx[e][][]\},\label{eq:tttttt}
\end{equation}
where $\xx[e][][]\in\EV_{\xx[o][][]}$, with $\EV_{\xx[o][][]}=\{\expt[\xx[o][][]][\dm[g]]:\dm[g]\in\SPd\}$
and $\SPd$ is the Liouville space for all density matrices; and we
use the notation $\expt[\hat{A}][\dm[g]]=\tr(\hat{A}\dm[g])$. The
function $\rho(\kk[c][][],\xx[e][][])$ provides the formal solution
to $\Ham$ for arbitrary values of $\kk[c][][]$ and $\xx[c][][]$.
Our main strategy is to introduce a trial density matrix using the
OET ansatz
\begin{equation}
\tilde{\rho}(\kk[c][][],\xx[e][][],\kk[e][][])=\Norm\mathcal{P}(\kk[c][][],\xx[e][][])\rho_{\kk[o][][]}(\kk[e][][])\mathcal{P}(\kk[c][][],\xx[e][][]),\label{eq:rhotilde}
\end{equation}
where $\Norm$ is the normalization, $\kk[e][][]\in\EVk$ with $\EVk=\{\expt[\kk[o][][]][\dm[g]]:\dm[g]\in\SPd\}$,
$\rho_{\kk[o][][]}(\kk[e][][])=\Norm'\exp(\kk[f][][]\kk[o][][])$
satisfying $\expt[\kk[o][][]][\rho_{\kk[o][][]}(\kk[e][][])]=\kk[e][][]$,
where $\Norm'$ is the normalization and $\kk[f][][]\in\mathbb{R}$.
Eq. \ref{eq:rhotilde} must satisfy the on-shell condition: for any
$\kk[c][][]$ there is a $\kk[e][][\Saddlep]\in\EVk$ such that $\tilde{\rho}(\kk[c][][],\xx[e][][],\kk[e][][\Saddlep])=\rho(\kk[c][][],\xx[e][][])$.
We can solve for $\mathcal{P}(\kk[c][][],\xx[e][][])$ using the on-shell
condition
\begin{equation}
\mathcal{P}(\kk[c][][],\xx[e][][])=\frac{1}{\sqrt{\EVxtoDMkg}}\left(\sqrt{\EVxtoDMkg}\rho(\kk[c][][],\xx[e][][])\sqrt{\EVxtoDMkg}\right)^{1/2}\frac{1}{\sqrt{\EVxtoDMkg}},\label{eq:Pkk-1-1-1-1}
\end{equation}
where $\EVxtoDMkg=\rho_{\kk[o][][]}(\kk[e][][\Saddlep])$. Finally,
the ground state energy can be constructed as
\begin{align}
\GSe(\RPcs) & =\lim_{\beta\rightarrow\infty}\min_{\kk[e][][]\in\EVk,\xx[e][][]\in\EVx}\expt[\Ham][\tilde{\rho}(\kk[c][][],\xx[e][][],\kk[e][][])].\label{eq:GroundStateEnergy-1-1-1-1-1}
\end{align}
It is useful to introduce the map $\Upsilon(\kk[c][][],\xx[e][][],\kk[e][][])=(\expt[\kk[o][][]][\tilde{\rho}(\kk[c][][],\xx[e][][],\kk[e][][])],\expt[\xx[o][][]][\tilde{\rho}(\kk[c][][],\xx[e][][],\kk[e][][])])$,
which is the essential quantity needed to execute the theory. Our
formalism has recast the exact solution of the Hamiltonian to a form
which will prove to be amenable to approximations. 

We now introduce the key approximation scheme: the central point expansion
(CPE). The CPE amounts to choosing an appropriate $\kk[e][][\Saddlep]$
and Taylor series expanding $\tilde{\rho}(\kk[c][][],\xx[e][][],\kk[e][][])$
in $\kk[c][][]$ and $\kk[e][][]$ about some \textit{central point}.
Here we choose the central point $\dm[g]_{\Center}\equiv\Norm\hat{1}$,
where $\Norm$ is the normalization, which yields $(\kk[e][\Center][],\xx[e][\Center][])=(\expt[\kk[o][][]][\dm[g]_{\Center}],\expt[\xx[o][][]][\dm[g]_{\Center}])$,
and we choose $\kk[e][][\Saddlep]$ such that $\mathcal{P}(\kk[c][][],\xx[e][\Center][])=1$
within our approximation. Expanding $\mathcal{P}(\kk[c][][],\xx[e][][])$
to zeroth order in $\kk[c][][]$ and $\rho_{\kk[o][][]}(\kk[e][][])$
to first order in $\kk[e][][]$ about $\kk[e][\Center][]$, we find
$\kk[e][][\Saddlep]=\kk[e][\Center][]$ and we have
\begin{align}
 & \mathcal{P}(\kk[c][][],\xx[e][][])\approx\mathcal{P}(0,\xx[e][][])=\sqrt{\rho_{\xx[o][][]}(\xx[e][][])\rho_{\xx[o][][]}^{-1}(\xx[e][\Center][])},\label{eq:Pkk-1-1-1-1-1-1}\\
 & \rho_{\kk[o][][]}(\kk[e][][])\approx\rho_{\kk[o][][]}(\kk[e][\Center][])(1+\aver[\Delta\hat{K}][\Delta\hat{K}][\rho_{\hat{X}}\left(X_{\Center}\right)]^{-1}\Delta\kk[o][][]\Delta\kk[e][][]),
\end{align}
where $\Delta\kk[o][][]=\kk[o][][]-\kk[e][\Center][]\hat{1}$, $\Delta\kk[e][][]=\kk[e][][]-\kk[e][\Center][]$,
and $\aver=\text{Tr}(\hat{A}\sqrt{\hat{\rho}}\hat{B}\sqrt{\hat{\rho}})$,
where the latter is known as the \textit{symmetric correlator}\cite{supplementary}.
To evaluate the ground state properties we only need to evaluate $\Delta\kk[o][][]$
and $\Delta\xx[o][][]$ under the CPE approximated $\tilde{\rho}(\kk[c][][],\xx[e][][],\kk[e][][])$,
denoted $\bar{\rho}$ for brevity
\begin{align}
 & \expt[\Delta\kk[o][][]][\bar{\rho}]=\lambda\Big(\expt[\Delta\hat{K}][\rho_{\hat{X}}\left(X\right)]+Z(\Delta\xx[e][][])\Delta K\Big),\\
 & \expt[\Delta\xx[o][][]][\bar{\rho}]=\lambda\Big(\Delta\xx[e][][]+\frac{\aver[\Delta\hat{X}][\Delta\hat{K}][\rho_{\hat{X}}\left(X\right)]}{\aver[\Delta\hat{K}][\Delta\hat{K}][\rho_{\hat{X}}\left(X_{\Center}\right)]}\Delta K\Big),\\
 & \lambda=\Big(1+\expt[\Delta\hat{K}][\rho_{\hat{X}}\left(X\right)]\aver[\Delta\hat{K}][\Delta\hat{K}][\rho_{\hat{X}}\left(X_{\Center}\right)]^{-1}\Delta K\Big)^{-1},\\
 & Z(\Delta\xx[e][][])=\aver[\Delta\hat{K}][\Delta\hat{K}][\rho_{\hat{X}}\left(X\right)]\aver[\Delta\hat{K}][\Delta\hat{K}][\rho_{\hat{X}}\left(X_{\Center}\right)]^{-1}.
\end{align}
The preceding expectation values approximate the map $\Upsilon(\kk[c][][],\xx[e][][],\kk[e][][])$,
and given that $\kk[c][][]=0$ within the CPE, we use a distinct symbol
$\bar{\Upsilon}(\xx[e][][],\kk[e][][])=(\expt[\kk[o][][]][\bar{\rho}],\expt[\xx[o][][]][\bar{\rho}])$. 

For a number of important Hamiltonians, including the Hubbard model
and its generalizations, we notice that $\expt[\Delta\hat{K}][\rho_{\hat{X}}\left(X\right)]=0$,
which implies that $\aver[\Delta\hat{X}][\Delta\hat{K}][\rho_{\hat{X}}\left(X\right)]=0$,
and we refer to this scenario as the \textit{orthogonal response condition}
(ORC)\cite{supplementary}. For Hamiltonians with a given partition
that satisfy the ORC, the CPE satisfies the exact condition $\bar{\Upsilon}\left(\Delta\kk[e][][],0\right)=(\Delta\kk[e][][],0)$,
and has the form $\bar{\Upsilon}\left(\Delta\kk[e][][],\Delta\xx[e][][]\right)=(Z(\Delta\xx[e][][])\Delta\kk[e][][],\Delta\xx[e][][])$;
all subsequent discussions of the CPE will presume the ORC. The CPE
will provide a reliable solution for $\Delta\xx[e][][]\ll\Delta\kk[e][][]$
and may provide reasonable solutions for $\Delta\xx[e][][]\approx\Delta\kk[e][][]$. 

Though the CPE has a non-perturbative structure in $\xx[e][][]$,
in addition to the favorable characteristics outlined above, it does
not have the correct second order expansion coefficient in $\Delta\xx[e][][]$.
Therefore, we introduce the \textit{renormalized central point expansion}
(RCPE)\cite{supplementary}, which maintains the form of $\bar{\Upsilon}$
but replaces $Z\rightarrow\pcorr(Z)$. Here we introduce perhaps the
simplest scheme where $\pcorr(Z)=\gamma_{0}Z^{\gamma_{1}}+(1-\gamma_{0})Z^{\gamma_{2}}$
and $\gamma_{1},\gamma_{2}$ are chosen from asymptotic analysis while
$\gamma_{0}$ is chosen to reproduce perturbation theory to second
order. It should be emphasized that $\pcorr$ has no free parameters.

The $\KKF$ formalism takes $\kk[c][][]\kk[o][][]$ as internal and
$\xx[c][][]\xx[o][][]$ as external, as previously defined. Alternatively,
we can invert internal and external to create a \textit{dual formulation},
which we refer to as the $\XXF$ formulation; and this can be obtained
by the substitutions 
\begin{align}
\KKF\leftrightarrow\XXF, &  & \kk[c][][]\leftrightarrow\xx[c][][], &  & \kk[e][][]\leftrightarrow\xx[e][][], &  & \kk[o][][]\leftrightarrow\xx[o][][].\label{eq:dualtransformation-1}
\end{align}
All equations within the $\KKF$ formalism will have a correspondence
in $\XXF$\cite{supplementary}, and therefore a subscript of $\KKF$
or $\XXF$ will be introduced when necessary. The $\XXF$ formulation
provides an opposite viewpoint of the physics, and exploring both
$\KKF$ and $\XXF$ will provide a more robust description of the
solution as each formulation will reproduce the exact second order
expansion in the corresponding limit. There could be many schemes
to choose between $\KKF$ and $\XXF$, and the total energy is a natural
candidate. Here we explore both approaches, and simply use continuity
when switching is necessary. 
\begin{figure}
\includegraphics[width=0.9\columnwidth]{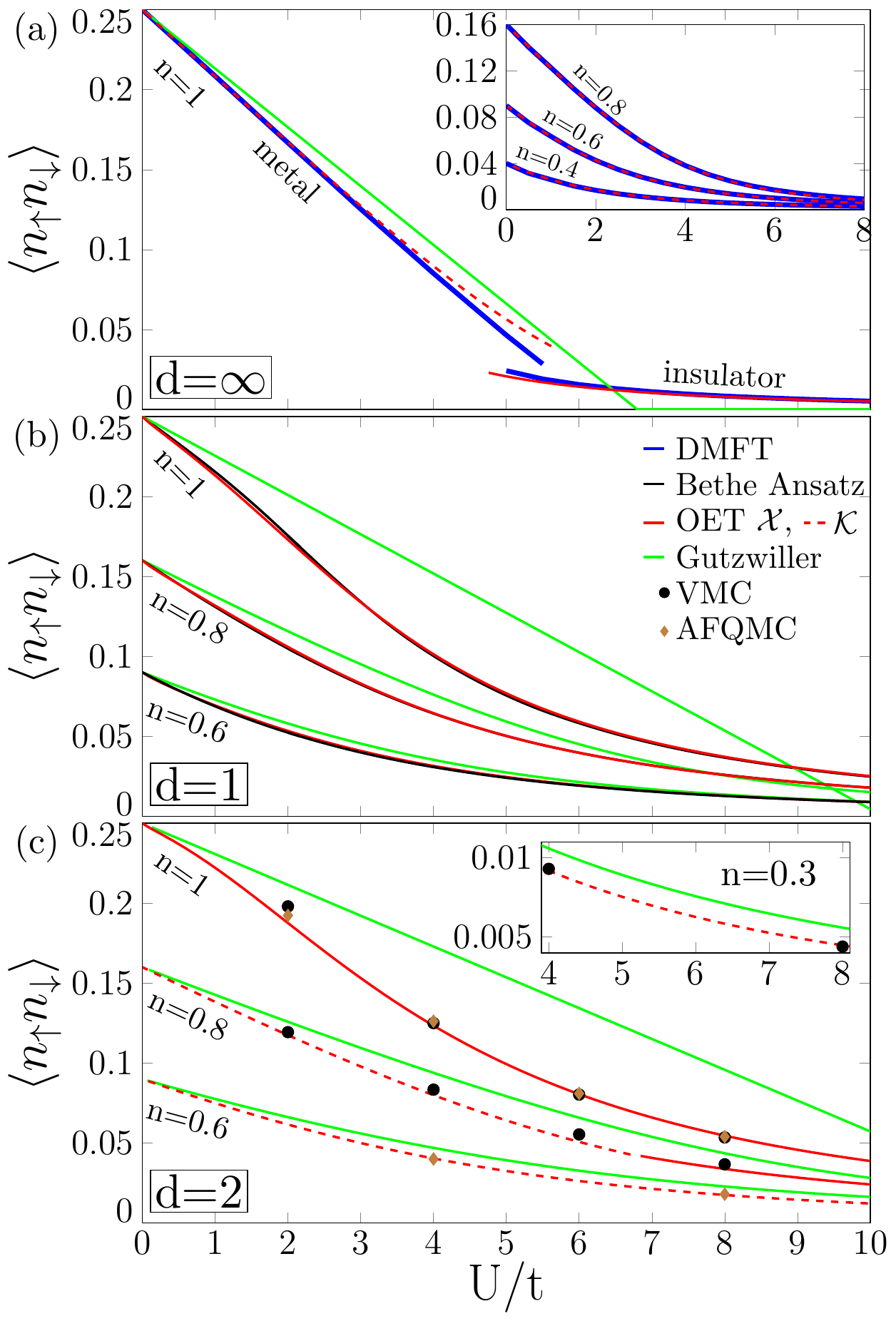}

\caption{\label{fig:fig1}Double occupancy for the Hubbard model in various
dimensions. (a) The $d$=$\infty$ Bethe lattice for various dopings,
solved within DMFT, GA, and OET. (b) The $d$=1 lattice, solved within
Bethe Ansatz, GA, and OET. (c) The $d$=2 square lattice solved with
GA, OET, and selected points using VMC and AFQMC \cite{Leblanc2015041041}. }
\end{figure}
Several simplifications were made in the above exposition of the OET
formalism and its approximations. Here we consider a more general
case applicable to many important Hamiltonians including Hubbard models.
We begin by considering a Hamiltonian partitioned into two parts,
where each portion is now resolved onto a set of commuting operators
\begin{align}
\Ham=\Ham_{\KK}+\Ham_{\XX}=\sum_{\KKt}\kk[gc]\kk[go]+\sum_{\XXt}\xx[gc]\xx[go],\label{eq:ham_comm}
\end{align}
where $[\kk[o][m][],\kk[o][m'][]]=[\xx[o][n][],\xx[o][n'][]]=0$.
A set of quantities $\{A_{i}\}$ (e.g. operators, expectation values,
etc) can be encoded as a vector, which is denoted as $\vec{A}=(A_{1},A_{2},\dots)$.
For example, we have $\Ham=\kk[Vc]\cdot\kk[Vo]+\xx[Vc]\cdot\xx[Vo]$.
We define the density matrix determined from $\vec{A}$ as $\rho_{\hat{\vec{A}}}(\vec{A})=\Norm\exp(\bm{\alpha}\cdot\hat{\vec{A}})$
satisfying $\expt[\hat{\vec{A}}][\rho_{\hat{\vec{A}}}(\vec{A})]=\vec{A}$,
where $\bm{\alpha}$ is a vector of real numbers, and the domain of
$\rho_{\hat{\vec{A}}}(\vec{A})$ is denoted $\EV_{\hat{\vec{A}}}=\{\expt[\hat{\vec{A}}]:\dm[g]\in\SPd\}$.
The ground state energy can then be written as
\begin{align}
\GSe(\RPc) & =\lim_{\beta\rightarrow\infty}\min_{\substack{\kk[Ve][][]\in\EV_{\kk[Vo]},\xx[Ve][][]\in\EV_{\xx[Vo]}}
}\expt[\Ham][\tilde{\rho}(\kk[Vc][][],\xx[Ve][][],\kk[Ve][][])].\label{eq:GroundStateEnergy-1-1-1-1-1-1}
\end{align}
We also define the map $\Upsilon(\kk[Vc][][],\xx[Ve][][],\kk[Ve][][])=(\expt[\kk[Vo][][]][\tilde{\rho}(\kk[Vc][][],\xx[Ve][][],\kk[Ve][][])],\expt[\xx[Vo][][]][\tilde{\rho}(\kk[Vc][][],\xx[Ve][][],\kk[Ve][][])])$,
which provides the complete solution to the Hamiltonian.

In order to implement the CPE in general, we need to find the independent
constraints between $\kk[Vo]$ and $\xx[Vo]$ (e.g. density), denoted
as $\cons[vo][]$, where $\cons[o]=\vec{\ConsK}_{i}\cdot\kk[Vo]=\vec{\ConsX}_{i}\cdot\xx[Vo]$.
The central point will be chosen as $\dm[g]_{\Center}=\rho_{\hat{\vec{C}}}(\vec{C})$
where $[\vec{C}]_{i}=\vec{\ConsK}_{i}\cdot\kk[Ve]=\vec{\ConsX}_{i}\cdot\xx[Ve]$. 

Here we test our formalism on the single band Hubbard model 
\begin{alignat}{1}
 & \Ham=\sum_{\KKp\sigma}\epsilon_{\KKp}\nn[o][\KKp\sigma]+N(U\hat{d}-\sum_{\sigma}\mu_{\sigma}\hat{n}_{\sigma}),\label{eq:Hubbard}
\end{alignat}
where $p$ labels a point in the first Brillouin Zone, $N$ is the
total number of sites in the lattice, $\hat{n}_{\sigma}=(1/N)\sum_{j}\hat{n}_{j\sigma}$
where $j$ labels a real space lattice point and $\hat{n}_{j\sigma}=\hat{a}_{j\sigma}^{\dagger}\hat{a}_{j\sigma}$,
$\mu_{\sigma}=\mu+h(\delta_{\uparrow\sigma}-\delta_{\downarrow\sigma})$,
and $\hat{d}=(1/N)\sum_{j}\nn[o][\XXi\uparrow]\nn[o][\XXi\downarrow]$.
To connect with Eq. \ref{eq:ham_comm}, we identify $\kk[Vo]=(\dots,\nn[o][\KKp\sigma],\dots)$,
$\xx[Vo]=(\hat{d},\nn[o][\uparrow],\nn[o][\downarrow])$, and $\cons[vo][]=(\nn[o][\uparrow],\nn[o][\downarrow])$.
For a given constraint $(\nn[e][\uparrow],\nn[e][\downarrow])$, we
parameterize $\kk[Ve]\in\EVk$ and $\xx[Ve]\in\EVx$ using $\Delta d=d-\nn[e][\uparrow]\nn[e][\downarrow]$
where $\Delta d\in[\Delta d_{min},\Delta d_{max}]$ and
\begin{align}
\Delta d_{min} & =-\min\left(\left(1-\nn[e][\uparrow]\right)\left(1-\nn[e][\downarrow]\right),\nn[e][\uparrow]\nn[e][\downarrow]\right),\label{eq:deltadmin}\\
\Delta d_{max} & =\min\left(\left(1-\nn[e][\uparrow]\right)\nn[e][\downarrow],\nn[e][\uparrow]\left(1-\nn[e][\downarrow]\right)\right),\label{eq:deltadmax}
\end{align}
and $\Delta\nn[e][\KKp\sigma]=\nn[e][\KKp\sigma]-\nn[e][\sigma]$,
where $\Delta\nn[e][\KKp\sigma]\in[-n_{\sigma},1-n_{\sigma}]$ and
the constraint requires $\sum_{\KKp}\Delta\nn[e][\KKp\sigma]=0$;
for brevity, we denote $\Delta\vec{n}=(\dots,\Delta\nn[e][\KKp\sigma],\dots)$.

We begin by presenting the CPE for both the $\KKF$ and $\XXF$ formalisms\cite{supplementary},
where the $\KKF$ formalism gives
\begin{align}
 & \bar{\Upsilon}_{\KKF}(\Delta\vec{n},\Delta d)=(\zz[knok][yes][(\KKp\sigma)]\Delta\vec{n},\Delta d),\hspace{1em}[\zz[knok][yes][]]_{\KKp\sigma}=\zz[knok][yesc][(\sigma)],\label{eq:KXcpe}\\
 & \zz[knok][yesc][(\sigma)]=\amp[k][yes](\Delta d)/\amp[k][yes][c](0),\hspace{1em}\rho_{\xx[Vo]}(\Delta d)={\textstyle \bigotimes\limits _{\XXi}}\rho_{\XXi}(\Delta d),\\
 & \amp[k][yes](\Delta d)=\aver[\fop[c][\XXi]][\fop[d][\XXi]][\rho_{\xx[Vo]}(\Delta d)]^{2},\\
 & \rho_{j}(\Delta d)=\textrm{diag}(p_{0},p_{\downarrow},p_{\uparrow},p_{2}),\hspace{1em}p_{2}=\nn[e][\uparrow]\nn[e][\downarrow]+\Delta d,\\
 & p_{0}={\textstyle \prod\nolimits _{\sigma}}(1-\nn[e][\sigma])+\Delta d,\hspace{1em}p_{\sigma}=(1-\nn[e][\bar{\sigma}])\nn[e][\sigma]-\Delta d,
\end{align}
and the $\XXF$ formulation gives
\begin{align}
 & \bar{\Upsilon}_{\XXF}(\Delta\vec{n},\Delta d)=(\Delta\vec{n},\zz[x][no]\Delta d),\\
 & \zz[x][no]=\amp[x][no](\Delta\vec{n})/\amp[x][no][c](\vec{0}),\hspace{1mm}\rho_{\kk[Vo]}(\Delta\vec{n})={\textstyle \bigotimes\limits _{\KKp\sigma}}\rho_{\KKp\sigma}(\Delta\nn[e][\KKp\sigma])\\
 & \amp[x][no](\Delta\vec{n})=(1/N^{4}){\textstyle \prod\nolimits _{\sigma}}|{\textstyle \sum\nolimits _{\KKp}}\aver[\fop[c][\KKp][\sigma]][\fop[d][\KKp][\sigma]][\rho_{\kk[Vo]}(\Delta\vec{n})]|^{2}\\
 & \rho_{\KKp\sigma}(\Delta\nn[e][\KKp\sigma])=\textrm{diag}(1-\nn[e][\sigma]-\Delta\nn[e][\KKp\sigma],\nn[e][\sigma]+\Delta\nn[e][\KKp\sigma])
\end{align}

The RCPE for the $\KKF$ formalism can be constructed as $\Upsilon_{\KKF}(\kk[Vc][][],\xx[Ve][][],\kk[Ve][][])=(\pcorrk(\kk[Vc],\vec{Z}{}_{\KKF})\Delta\vec{n},\Delta d)$
with $[\pcorrk(\kk[Vc],\vec{Z}{}_{\KKF})]_{\KKp\sigma}=\gamma_{0}(Z_{\KKF}^{(\sigma)})^{\gamma_{1}}+(1-\gamma_{0})(Z_{\KKF}^{(\sigma)})^{\gamma_{2}}$
and $\gamma_{1}=1$ and $\gamma_{2}=1/2$ \cite{supplementary}. Similarly,
for the $\XXF$ formalism we have $\Upsilon_{\XXF}(\xx[Vc][][],\kk[Ve][][],\xx[Ve][][])=(\Delta\vec{n},\pcorrx(\xx[Vc],Z{}_{\XXF})\Delta d)$,
where $\pcorrx(\xx[Vc],Z{}_{\XXF})=\gamma_{0}(Z_{\XXF}^{(\sigma)})^{\gamma_{1}}+(1-\gamma_{0})(Z_{\XXF}^{(\sigma)})^{\gamma_{2}}$
and $\gamma_{1}=1$ when there is no short range magnetic order (i.e.
paramagnetic state in $d$=$\infty$) while $\gamma_{1}=1/2$ for short
or long range antiferromagnetic order; and $\gamma_{2}=1/4$ in all
cases\cite{supplementary}. In both $\KKF$ and $\XXF$, $\gamma_{0}$
is uniquely determined from perturbation theory, such that there are
no free parameters within the RCPE. 

It should be noted that within the CPE (i.e. without renormalization),
the classic Gutzwiller approximation (GA)\cite{Gutzwiller1963159,Gutzwiller1964923,Gutzwiller19651726,Metzner19887382,Lanata2015011008}
to the Hubbard model is rigorously recovered, providing a qualitative
description of the Fermi liquid phase; similar to slave Bosons\cite{Kotliar19861362,Lechermann2007155102,Piefke2018125154}
and DMET\cite{Knizia2012186404,Ayral2017235139,Lee2019115129}. Therefore,
the RCPE in the $\KKF$ formulation is a clear improvement of Gutzwiller
and related approximations. Alternatively, the $\XXF$ formulation
within the RCPE will be shown to provide a robust description of the
Luttinger liquid and the Mott insulator, and we are not aware of a
corresponding result; though a related approach has been proposed
in the Baeriswyl wave function and its extensions\cite{Baeriswyl19860,Otsuka19921645,Baeriswyl20002033,Eichenberger2007180504,Eichenberger2009100510,Hetenyi2010115104}.
Furthermore, we note that the maps $\Upsilon_{\KKF},\Upsilon_{\XXF}$
directly provide a description of the physical space of all $(\expt[\Delta\hat{\vec{n}}][\dm[g]],\expt[\Delta\hat{d}][\dm[g]])$,
yielding a concrete approximation that resolves the N-representability
problem\cite{Mayer19551579,Coleman1963668,Garrod19641756,Davidson1969725,Ayers2005507,Mazziotti2012263002}
in this class of Hamiltonians. Therefore, OET provides an alternative
viewpoint to this problem, which is of strong interest in the field
of quantum chemistry and solid state physics\cite{Muller1984446,Zhao20042095,Piris2017063002,Mitxelena20190,Lathiotakis2014032511,Theophilou20184072,Blochl2011205101,Blochl2013205139,Kamil2016085141,Schade2018245131}.

\begin{figure}
\includegraphics[width=1\linewidth]{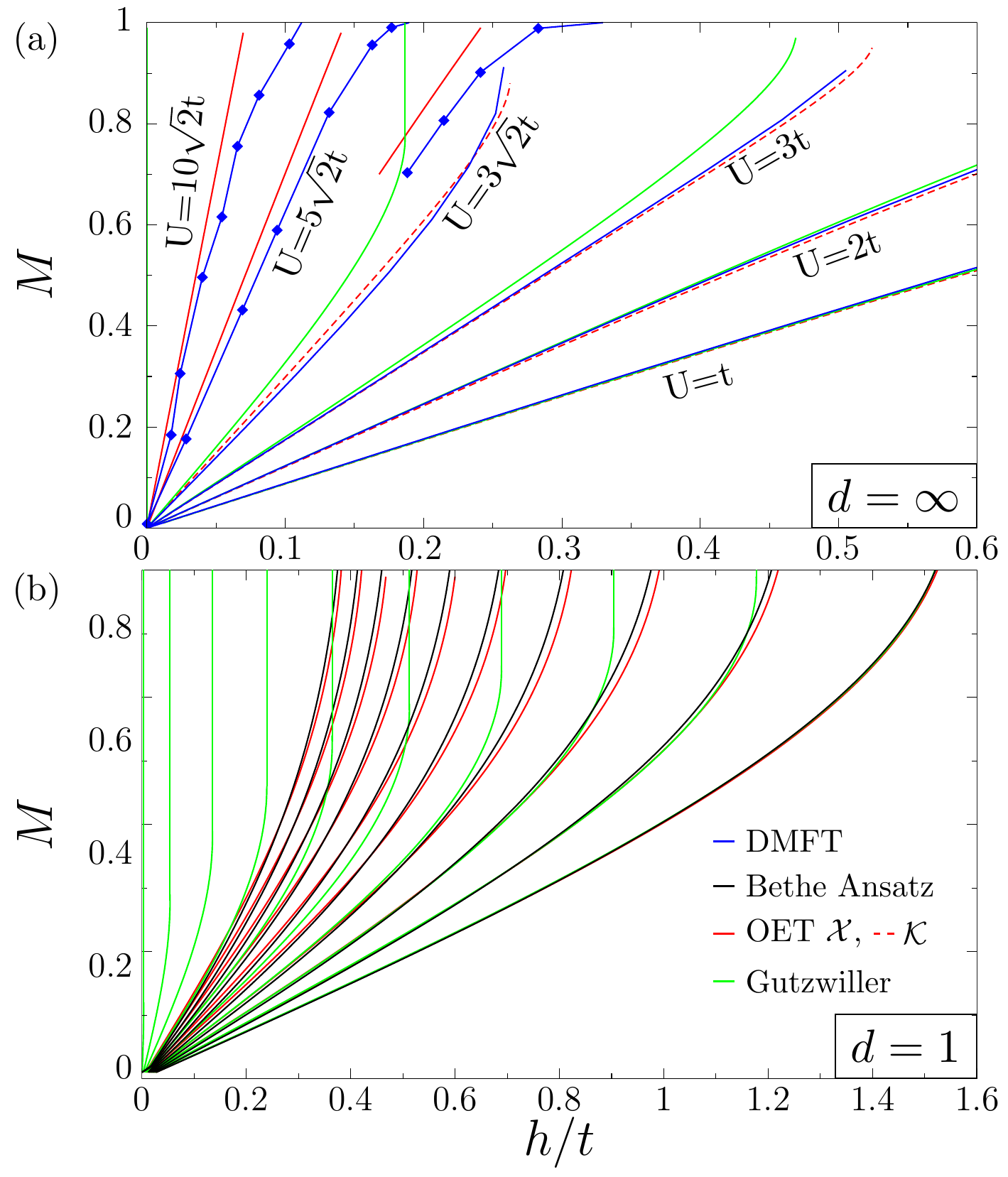}\caption{\label{fig:fig2}Magnetization $M$ vs. applied field $h$ for the
Hubbard model in $d$=$\infty$ and $d$=$1$. (a) The $d$=$\infty$ Bethe
lattice solved within DMFT (insulating results from Ref. \cite{Georges199613}),
GA, and OET. (b) The $d$=$1$ lattice solved within the Bethe Ansatz,
GA, and OET for $U/t=1,..,10$ (right to left).}
\end{figure}

We now apply OET for the Hubbard model in $d$=1, 2, $\infty$ over a broad
range of $t$, $U$, and density. In addition to comparing with exact
or state-of-the-art methods, we will also compare with the Gutzwiller
approximation given that it is an efficient approach. We begin with
$d$=$\infty$ at half-filling, where we examine the double occupancy
as a function of $U/t$ (see Figure \ref{fig:fig1}a). The DMFT results
are formally exact for $d$=$\infty$, and numerical renormalization
group\cite{Wilson1983583} is used to solve the DMFT impurity problem\cite{Sakai1994307,Bulla1999136,Zitko2009085106,Georges199613}
as implemented in the ``NRG Ljubljana'' code\cite{Zitkonrg}. The
DMFT results are denoted by blue lines, while the Gutzwiller results
are in green,. Gutzwiller yields a qualitative description of the
metallic phase, whereas the insulator is improperly described as a
collection of atoms. The OET results are given in red, with a dashed
line for $\KKF$ and solid for $\XXF$, showing favorable agreement
with DMFT in both the metallic and insulating regimes. The inset illustrates
OET for doped cases, showing excellent agreement with DMFT. 

We now turn to $d$=$1$ and the $d$=$2$ square lattice with nearest neighbor
hopping, where we examine the double occupancy versus $U/t$ for various
densities (see Figure \ref{fig:fig1}b, c). In one dimension, we compare
to the exact Bethe Ansatz solution\cite{Lieb19681445,Lieb20031},
while in two dimensions we compare to variational quantum Monte-Carlo
(VMC) and Auxiliary Field Quantum Monte-Carlo \cite{Leblanc2015041041}.
In one dimension (Figure \ref{fig:fig1}b), the OET $\XXF$ formulation
shows remarkable agreement with the BA, both at half filling and for
doped cases, and the $\KKF$ formulation is found not to be necessary.
In two dimensions, OET is also in good agreement with the VMC and
AFQMC results, both at half filling and for the doped cases; and here
continuity is used to switch between the $\KKF$ and $\XXF$ formulations
(Figure \ref{fig:fig1}c).
\begin{figure}
\includegraphics[width=1\linewidth]{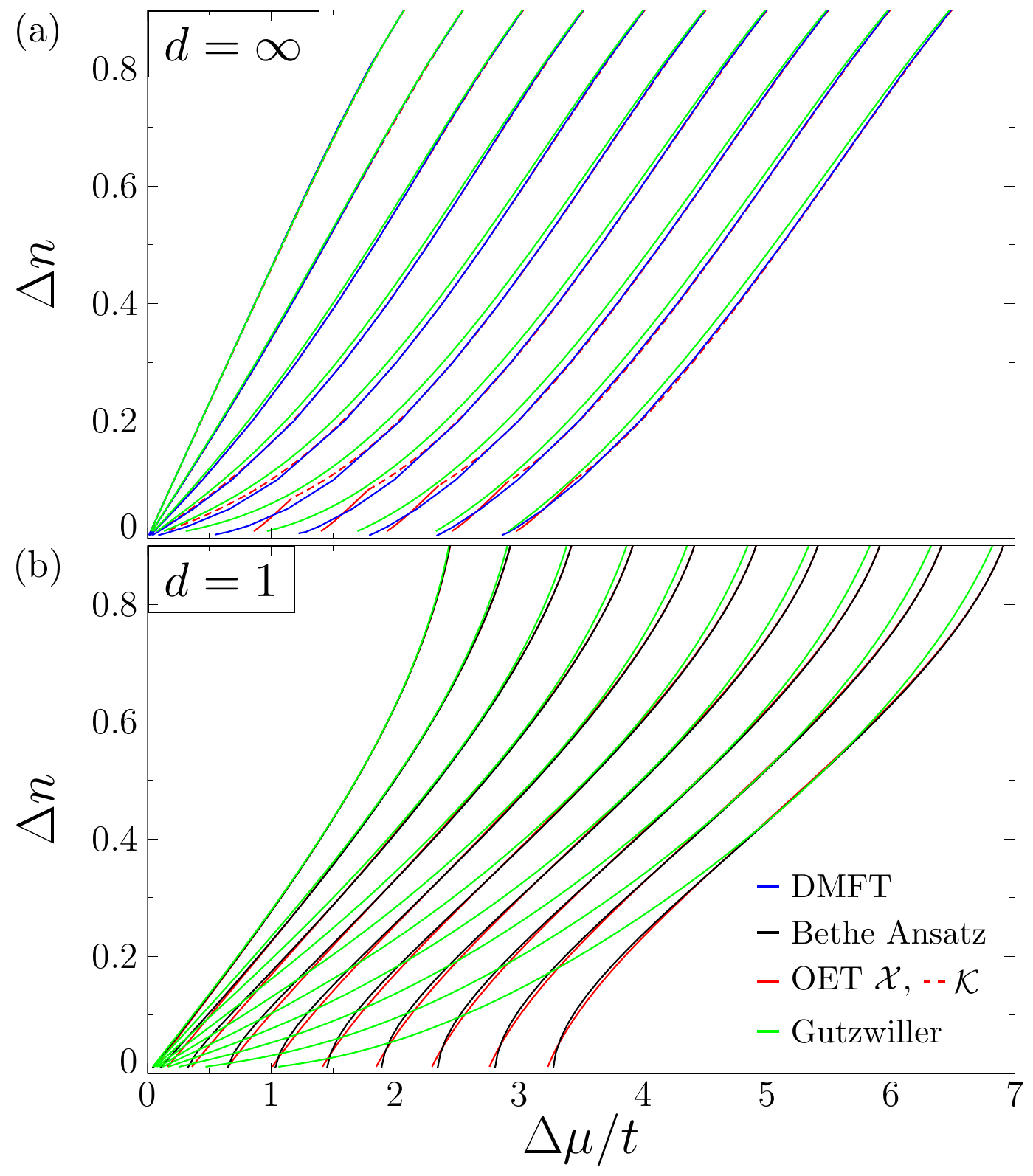}\caption{\label{fig:3}The density ($\Delta n=n-1$) as a function of chemical
potential ($\Delta\mu=\mu-U/2$) for the Hubbard model in $d$=$\infty$
and $d$=$1$ for $U/t=1,..,10$ (left to right). (a) The $d$=$\infty$
Bethe lattice solved within DMFT, GA, and OET. (b) The $d$=$1$ lattice
solved with the Bethe Ansatz, GA, and OET.}
\end{figure}

The magnetization under applied magnetic field for $d$=$\infty$ is
accurately captured using OET, even reasonably describing the coexistence
region between metal and insulator (see Figure \ref{fig:fig2}a).
For $d$=$1$, OET has excellent agreement over all parameters. In both
$d$=$\infty$ and $d$=$1$, Gutzwiller discontinuously polarizes for sufficiently
large $U$. The density as a function of the chemical potential for
$U/t=1,\dots,10$ is computed in $d$=$\infty$ and $d$=$1$ (Figure \ref{fig:3}).
For $d$=$\infty$, the system opens a gap at a finite $U$, and the
$\KKF$ and $\XXF$ ansatz can reasonably capture this behavior (Figure
\ref{fig:3}a). For $d$=$1$ , it is well known that any finite $U$
opens a gap, and this property is captured using the $\XXF$ formulation,
yielding favorable agreement over all parameters (Figure \ref{fig:3}b).
Results for $d$=$2$ can be found in Ref. \cite{supplementary}, Figure
1.

In summary, we have developed an exact formalism (i.e. OET) and a
generic approximation scheme (i.e. RCPE) for solving the ground state
of quantum Hamiltonians. Our approach is proven to be efficient and
globally robust for the one band Hubbard model in $d$=1, 2, $\infty$.
The success of our approach is based on four key ideas: the exact
OET construction, a non-perturbative form given by the CPE, a perturbative
correction given by the RCPE, and the combination of the dual forms
$\KKF$ and $\XXF$. Our approach can be straightforwardly applied
to a multitude of important quantum Hamiltonians. Furthermore, our
entire formalism can be generalized to finite temperature, and this
will be presented in a forthcoming paper. Finally, OET can straightforwardly
be combined with DFT, similar to DFT+DMFT\cite{Kotliar2006865} and
DFT+Gutzwiller\cite{Deng2009075114}, resulting in a highly efficient
first-principles approach to the thermodynamics of strongly correlated
materials in addition to molecules.

This work was supported by the grant DE-SC0016507 funded by the U.S.
Department of Energy, Office of Science. This research used resources
of the National Energy Research Scientific Computing Center, a DOE
Office of Science User Facility supported by the Office of Science
of the U.S. Department of Energy under Contract No. DE-AC02-05CH11231.
We thank R. Zitko for assistance with NRG DMFT calculations. We thank
G. Kotliar and K. Haule for useful discussions.

%

\end{document}